# Muon Collider Progress: Accelerators

Michael S. Zisman
*LBNL, Berkeley, CA 94720, USA*

A muon collider would be a powerful tool for exploring the energy-frontier with leptons, and would complement the studies now under way at the LHC. Such a device would offer several important benefits. Muons, like electrons, are point particles so the full center-of-mass energy is available for particle production. Moreover, on account of their higher mass, muons give rise to very little synchrotron radiation and produce very little beamstrahlung. The first feature permits the use of a circular collider that can make efficient use of the expensive rf system and whose footprint is compatible with an existing laboratory site. The second feature leads to a relatively narrow energy spread at the collision point. Designing an accelerator complex for a muon collider is a challenging task. Firstly, the muons are produced as a tertiary beam, so a high-power proton beam and a target that can withstand it are needed to provide the required luminosity of ~$1 \times 10^{34}$ cm$^{-2}$s$^{-1}$. Secondly, the beam is initially produced with a large 6D phase space, which necessitates a scheme for reducing the muon beam emittance ("cooling"). Finally, the muon has a short lifetime so all beam manipulations must be done very rapidly. The Muon Accelerator Program, led by Fermilab and including a number of U.S. national laboratories and universities, has undertaken design and R&D activities aimed toward the eventual construction of a muon collider. Design features of such a facility and the supporting R&D program are described.

## 1. Introduction

A muon collider would be a powerful tool in the experimentalist's arsenal, and design efforts and performance evaluations for such a facility have been ongoing for more than a decade. Initially, the work was carried out by the U.S. Neutrino Factory and Muon Collider Collaboration [1]. More recently, Fermilab's Muon Collider Task Force [2] joined the effort, with the program coordinated by the combined leadership of the two groups. In the past year, the two groups have been merged to form the Muon Accelerator Program (MAP, see Section 5.1). Recent interest by Fermilab management has spurred increased emphasis on developing and understanding the Muon Collider (MC) design. In what follows, we first describe the advantages and challenges of a Muon Collider and then discuss the subsystems that would comprise it. Finally, we briefly discuss the MAP and the R&D issues that it is studying.

## 2. Muon Accelerator Advantages and Challenges

### 2.1. Advantages

There are important particle physics questions at both the energy and intensity frontiers that can be addressed by muon accelerators.

At the energy frontier, the fact that the muon is a point particle means that the full beam energy is available for particle production. Due to its having a much larger mass than the electron, the muon emits almost no synchrotron radiation during acceleration. Thus, it is practical to utilize circular accelerators, which make more efficient use of rf equipment and have a smaller footprint[1] than do electron machines. Moreover, their larger mass means that muons in collision do not experience significant beamstrahlung, which results (see Fig. 1) in a narrower energy spread at the interaction point compared with an electron-positron collider.

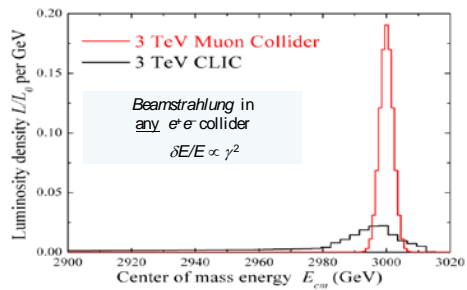

Figure 1: Comparison of calculated luminosity spectra at a 3 TeV Muon Collider and a 3 TeV CLIC e⁻e⁺ collider.

---

[1] For example, a Muon Collider would fit easily on the Fermilab site and would provide a world-class science program there.

At the intensity frontier, a beam of neutrinos from muon decay can serve as a Neutrino Factory. Such a neutrino beam, which contains an equal mixture of electron neutrinos and muon anti-neutrinos (or vice versa, depending on the charge of the decaying muons) gives high energy $\nu_e$ above the tau threshold. The $\nu_e \rightarrow \nu_\mu$ oscillations give rise to easily detectable "wrong-sign" muons. A Neutrino Factory has been shown [3] to provide unmatched sensitivity for studies of CP violation, the neutrino mass hierarchy, and unitarity in the neutrino sector.

## 2.2.    Challenges

There are two main challenges associated with using intense muon beams in an accelerator or collider:

1. Muons are created as a tertiary beam ($p \rightarrow \pi \rightarrow \mu$), and
2. Muons have a short lifetime (only 2.2 µs at rest).

The first challenge results in muons being produced at a low rate, so a multi-MW proton beam, along with a target capable of withstanding such intensity, is required. Furthermore, this production process leads to a muon beam with a large transverse phase space and a large energy spread. Handling such a beam requires a scheme for reducing its emittance ("cooling") and, even so, a high acceptance is needed for the acceleration system.

The second challenge puts a premium on rapid beam manipulations, in particular the cooling process, and demands that high-gradient rf cavities be employed in the cooling and acceleration systems. While several techniques for cooling are available for stable beams, only the presently untested process of ionization cooling is fast enough to be used for muons. Another consequence of the short muon lifetime is that electrons from muon decay represent a substantial heat load in collider ring magnets and a potentially large source of background in the detectors.

## 3.  Muon Collider Overview

As illustrated in Fig. 2, a Muon Collider consists of the following systems:

- A proton driver to provide the primary beam to the production target; the HARP experiment [4] addressed this aspect of the facility.
- A target, capture, and decay section, where pions are created, captured, and allowed to decay to muons; the MERIT experiment [5] addressed this aspect of the facility.
- A bunching and phase rotation section, where the beam is rotated in longitudinal phase space to reduce its energy spread.
- A cooling section, where the beam transverse and longitudinal phase space are reduced; the MICE experiment [6] addresses some aspects of this portion of the facility.
- An acceleration section, where the cooled muons are accelerated from ~10 MeV to ~1 TeV in a series of accelerators including a linac, recirculating linear accelerators (RLAs), fixed-field alternating gradient (FFAG) rings, and rapid-cycling synchrotrons (RCSs); the EMMA experiment [7] addresses the FFAG portion of the facility.
- A collider ring, where the muons are stored and collide at two interaction points for about 1000 turns.

Many of the low-energy systems listed above are similar or identical to those required for a Neutrino Factory, so much of the R&D for these two facilities can be done in common.

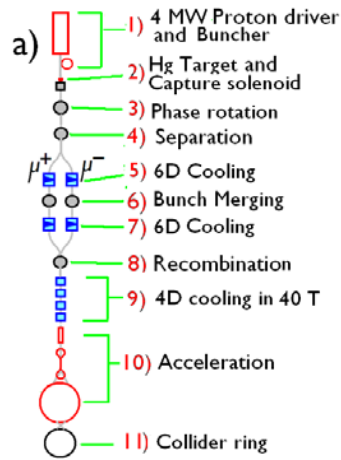

Figure 2: Schematic drawing of Muon Collider showing its various required systems.

Table 1 summarizes some example parameters for two of the Muon Collider scenarios presently under study [8]. The proton driver beam power estimates are based on assumed transmission efficiencies and may change as the designs are refined.

Table 1: Muon Collider example parameters.

| | | |
|---|---|---|
| $E_{c.m.}$ [TeV] | 1.5 | 3 |
| $L$ [$10^{34}$ cm$^{-2}$s$^{-1}$] | 1 | 4 |
| Beam-beam tune shift | 0.087 | 0.087 |
| Muons per bunch [$10^{12}$] | 2 | 2 |
| Total muon beam power [MW] | 9 | 15 |
| Avg. ring bend field [T] | 6.0 | 8.4 |
| $C$ [km] | 2.6 | 4.5 |
| $\beta^*$ [mm] | 10 | 5 |
| $\sigma_l$ [mm] | 10 | 5 |
| $\sigma_p$ [%] | 0.1 | 0.1 |
| $f_{rf}$ [MHz] | 805 | 805 |
| $V_{rf}$ [MV] | 20 | 230 |
| Repetition rate [Hz] | 15 | 12 |
| Proton driver power [MW] | ~4 | ~4 |
| $\varepsilon_\perp$ [μm] | 25 | 25 |
| $\varepsilon_{||}$ [mm] | 72 | 72 |

# 4. Muon Collider Description

## 4.1. Target and Capture

The baseline target for the Muon Collider is a free Hg jet. Simulated production rates from this target form the basis of the proton driver power requirements listed in Table 1. The capture section starts with a 20-T hybrid solenoid followed by a tapered solenoidal channel that brings the field down to the 1.5 T level used in the downstream systems. The high-radiation environment in which the 20-T solenoid is located makes this a particularly challenging region to design, and extra shielding (see Fig. 3) has recently been added to reduce the heat loads on the magnets closest to the target. Capture of low-energy pions (~100–350 MeV) is optimal for the downstream systems.

Based on muon production rates estimated with MARS15 [9], the optimum proton energy for producing pions in the energy range of interest is about 8 GeV. We take this as our design energy. Previous simulations [10] have also shown a preference for short bunches, ideally with $\sigma_l \sim 1$ ns. Bunch lengths up to 3 ns (rms) reduce the intensity by only 10%, and are deemed acceptable. For the short bunches we require, the minimum repetition rate is likely limited by the space-charge tune shift in the compressor ring. In order to reach the intensities required at a repetition rate of only 12–15 Hz, the current concept is to extract lower intensity bunches from the compressor ring and combine them at the target by transporting them through "delay lines," as illustrated schematically in Fig. 4.

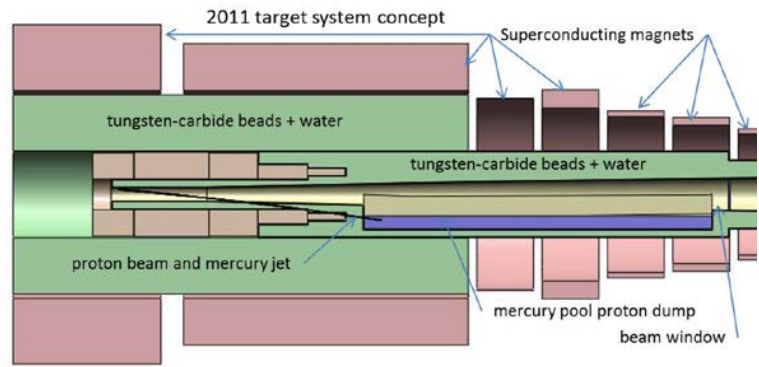

Figure 3: Muon Collider target area. Water-cooled tungsten carbide beads serve as shielding material.

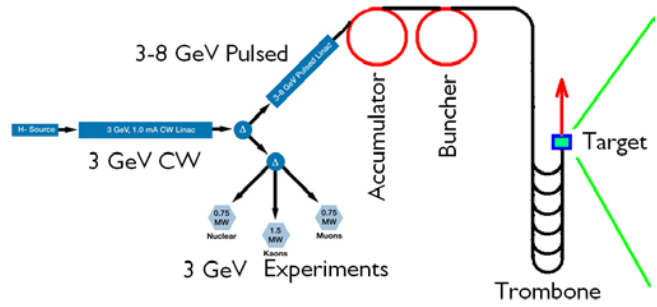

Figure 4: Illustration of "trombone" bunch combiner concept for the Fermilab Project X proton driver scenario.

## 4.2.    Bunching and Phase Rotation

The beam from the target is unsuitable for transmission through the downstream accelerators and must be conditioned to make it acceptable. After a drift section to create a correlation between momentum and position, the bunching and phase rotation system accomplishes this using rf frequencies that decrease along the channel. Figure 5 illustrates the effect on the muon beam and the distribution of frequencies utilized.

For a Muon Collider we desire the shortest possible bunch train, as our ultimate goal is to compress each train into a single bunch. This is done with a "6D" bunch-merging system, that is, a system that does some transverse merging and some longitudinal merging, as illustrated in Fig. 6.

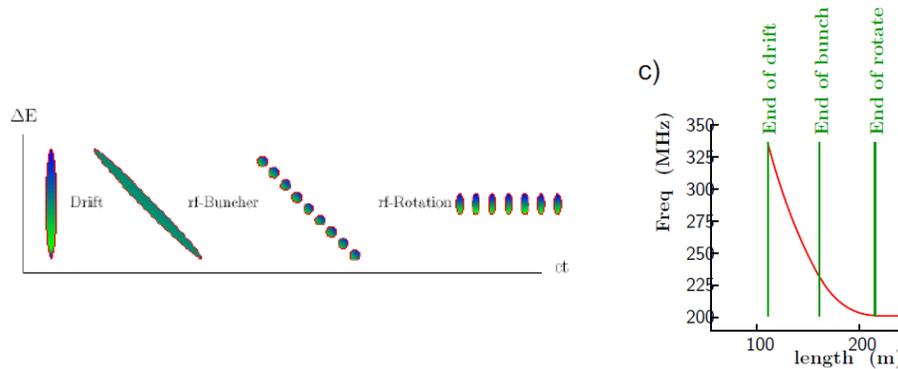

Figure 5: (left) Phase rotation and bunching of initial bunch from the target; (right) distribution of rf frequencies needed.

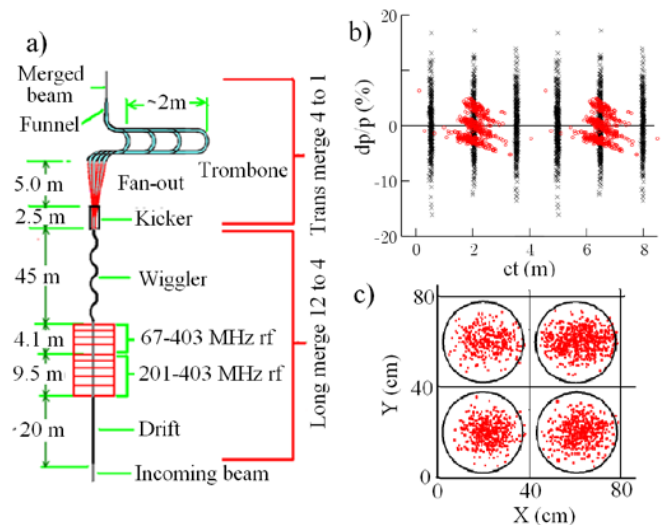

Figure 6: (a) Schematic of merging scenario; (b) longitudinal merge; (c) preparation for transverse merge.

### 4.3. Cooling

As noted earlier, cooling a muon beam requires the application of ionization cooling. The transverse cooling process is conceptually simple, as indicated in Fig. 7. The muons lose energy in all three dimensions via ionization of an absorber material, reducing $p_x$, $p_y$, and $p_z$. Energy gain in rf cavities restores only $p_z$. Repeating this many times reduces $p_{x,y}/p_z$ and thus represents transverse cooling. The performance of a cooling channel suitable for a Neutrino Factory or the initial portion of a Muon Collider is illustrated in Fig. 8. Note that such a channel transmits both muon signs on opposite phases of the rf cycle. The actual hardware implementation of such a channel—where high power rf cavities are in close proximity to liquid-hydrogen absorbers—is complicated. Fortunately, means of dealing with the issues have been developed as part of the preparations for MICE [6].

Cooling in 6D requires an additional step—emittance exchange. The idea is to increase the energy loss of high-energy muons compared with that for low-energy muons. Conceptually, there are several ways to accomplish this. Dispersion can be introduced into the beam line and a wedge-shaped absorber can be employed to accomplish the differential energy loss. A variant is to use a continuous absorber in a dispersive region, such that the higher energy particles have a longer path length and thus lose more energy than lower energy particles, which follow a shorter path.

Several possible implementations of such a scheme, including a "Guggenheim" channel, a FOFO snake, and a helical cooling channel (HCC), are shown in Fig. 9. In the simulations to date, a pair of Guggenheim channels (one for each muon sign) has been assumed. Having an initial FOFO snake channel, which accommodates both signs, is an attractive possibility that will be explored in the future.

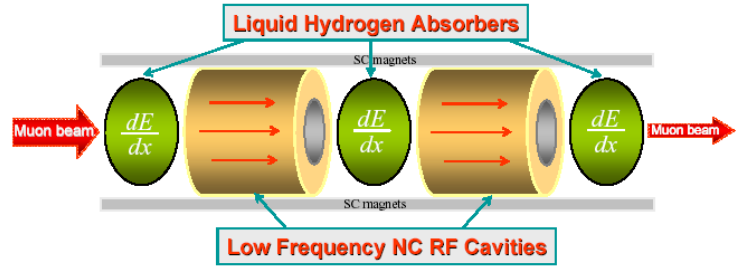

Figure 7: Schematic of 4D ionization cooling system.

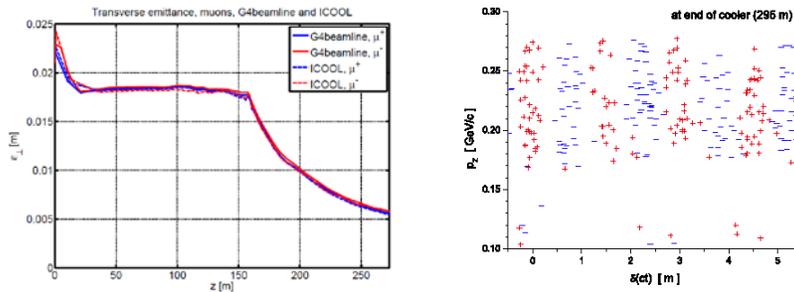

Figure 8: (left) Performance of initial 4D cooling channel; (right) interleaved bunches of positive (red) and negative (blue) muons at the exit of the cooling channel.

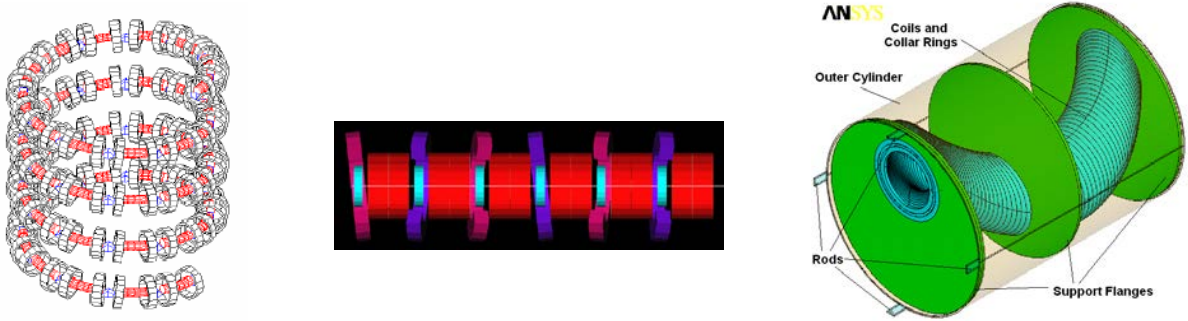

Figure 9: Possible implementations of 6D cooling. (left) Guggenheim channel; (center) FOFO snake; (right) HCC.

### 4.4.    Final Cooling

The final cooling of the muons to a normalized transverse emittance of 25 μm is without question the most challenging aspect of the facility. A concept that is presently under development [11] is illustrated in Fig. 10. In order to reach the design emittance, very strong (~30–50 T) solenoids are assumed [2]. Although a 45 T solenoid has been successfully built [12] at the National High Magnetic Field Laboratory, it is not a very practical device for an accelerator. Thus, this remains an R&D topic. Several groups are exploring the idea of using high-temperature superconductor (HTS) material for this task, as this is thought to be the most likely technology to work.

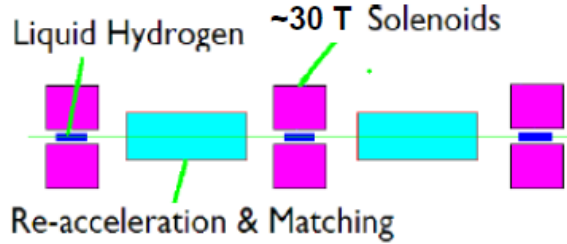

Figure 10: Schematic of Muon Collider final cooling scheme.

### 4.5.    Acceleration

The acceleration system for a Muon Collider must increase the beam energy by more than a factor of 100. To make efficient use of the acceleration hardware, this is accomplished in stages that are matched to particular energy ranges. The obvious drawback in this approach is that there are many distinct interface points that must be designed and optimized, but the alternative of having only one or a few stages is deemed to be neither practical nor cost-effective.

The low-energy portion of the system, which is expected to be common to that for a Neutrino Factory, is illustrated in Fig. 11. It comprises a linac, followed by a pair of RLAs and then a non-scaling FFAG ring. These devices all have a 30 mm transverse and 150 mm longitudinal acceptance and keep both muon signs [13].

The baseline scheme for the high-energy portion involves a pair [3] of RCSs that would be located in the Tevatron tunnel at Fermilab, the first accelerating from 25 to 400 GeV and the second from 400 to 750 GeV. To reach the desired repetition rate of 400 Hz, special magnets are required that use grain-oriented silicon steel laminations. For the second RCS, superconducting magnets are needed to keep the circumference compatible with the Tevatron tunnel, but these cannot be ramped quickly. The approach used to handle this is shown in Fig. 12. Fast-ramping conventional magnets (of the type used in the first RCS) are interleaved with the fixed-field superconducting magnets. At the beginning of the ramp, the conventional magnet fields oppose those of the superconducting magnets, and at the end of the ramp they reinforce them.

### 4.6.    Collider Ring

A lattice design for a 1.5 TeV Muon Collider has been developed [14] and is shown in Fig. 13. The bare lattice has adequate dynamic aperture and momentum acceptance. Studies with errors remain to be carried out. Work on developing a lattice for a 3 TeV collider is presently under way.

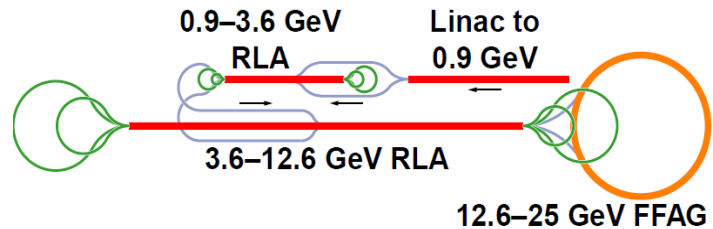

Figure 11: Low-energy Muon Collider acceleration scheme.

---

[2] There is no "hard edge" to the field requirement, but higher fields ease the requirements on other channel parameters.
[3] To reach the 1.5 TeV energy needed for a 3 TeV collider, another RCS would be required. This device would be bigger and would not fit in the Tevatron tunnel.

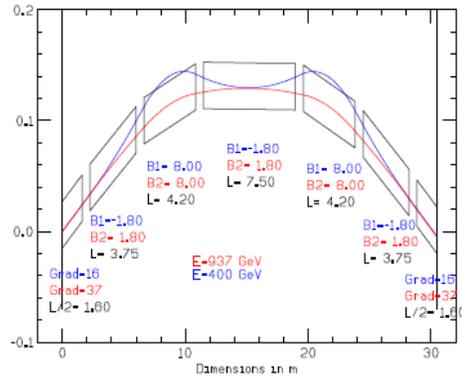

Figure 12: Higher-energy scheme for Muon Collider acceleration.

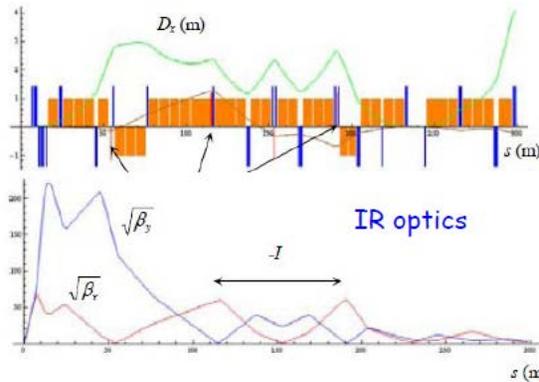

Figure 13: 1.5 TeV Muon Collider interaction region (upper) and optics (lower).

### 4.7. Machine-Detector Interface

As is the case for most modern colliders, the machine-detector interface [8] is of critical importance and thus represents a key design activity. This aspect of the design is what determines the ultimate physics capability of the facility, in the sense that it permits an assessment of both the expected backgrounds and the techniques to mitigate them. Recent work on this topic suggests that timing information can drastically reduce the anticipated backgrounds. A successful collider implementation requires that the detector and shielding be tightly integrated into the machine design. This requires a close collaboration between machine designers and the future users of the facility, which is now beginning to happen.

## 5. R&D Program

### 5.1. Muon Accelerator Program

A substantial R&D program is under way to validate Muon Collider design choices. The program has three components: simulations, technology development, and system tests. At present, Muon Collider R&D is primarily a U.S. enterprise, but it is hoped that the program will become international over time, as did the Neutrino Factory R&D activity [3]. As mentioned earlier, Muon Collider R&D activities are managed via the MAP.

MAP was set up by Fermilab at the request of the U.S. Department of Energy to deliver several things:

- a Design Feasibility Study (DFS) report on the Muon Collider that will include an assessment of feasibility and a cost range for such a facility
- a technology development program that will inform the DFS and allow down-selection of the key technologies
- participation, under the auspices of the IDS-NF, in preparing a Reference Design Report (RDR) for a Neutrino Factory
- participation in MICE

The aim—dependent on obtaining the requisite level of funding—is to complete the DFS in a 2016 time frame. In addition to the accelerator R&D program, it is expected that a parallel physics and detector study will be carried out during this period.

MAP has developed a Mission Statement to define its work:

> *The mission of the Muon Accelerator Program (MAP) is to develop and demonstrate the concepts and critical technologies required to produce, capture, condition, accelerate, and store intense beams of muons for Muon Colliders and Neutrino Factories. The goal of MAP is to deliver results that will permit the high-energy physics community to make an informed choice of the optimal path to a high-energy lepton collider and/or a next-generation neutrino beam facility. Coordination with the parallel Muon Collider Physics and Detector Study and with the International Design Study of a Neutrino Factory will ensure MAP responsiveness to physics requirements.*

### 5.2. R&D Issues

While the ongoing R&D program details cannot be included here, it is worthwhile to identify the main issues. For the simulation effort, the main tasks are to optimize the designs of the various subsystems and then to carry out end-to-end simulations of the entire facility in order to define its expected performance.

For the technology development, the key issue is to understand and improve the performance of normal-conducting rf cavities in an axial magnetic field. Other tasks include:

- development of low-frequency superconducting rf cavities
- development of high-field solenoids for final cooling
- development of fast ramping magnets for the RCS
- development of decay ring magnets that can withstand the mid-plane heat load from muon decay products

In terms of system tests, one such experiment, MERIT, has already been completed, and another, MICE, is under way. As part of the preparations for a Muon Collider, MAP will assess the need for a full-fledged 6D cooling experiment. If it is deemed necessary, a proposal for an experiment will be prepared. Actually carrying out such a major undertaking, however, is considered to be beyond the scope of the initial phase of MAP. If such an experiment is deemed necessary, it would be carried out in a subsequent phase of MAP, hopefully with international participation.

## 6. Summary

As we have seen in this report, the accelerator design for a Muon Collider is progressing well. A promising collider lattice has been developed, and all of the major subsystems of the collider have been simulated to some degree. What remains is to tie the subsystems together and carry out the end-to-end simulations and collective effects studies that provide a realistic assessment of overall performance.

The R&D toward a Muon Collider is also making steady progress. The MERIT experiment has established the ability of a mercury-jet target to tolerate more than 4 MW of proton beam. MICE is also moving forward. The muon beam line is commissioned and characterized, and all of the major cooling channel components are in production. We are looking forward to the initial measurements of ionization cooling in the next few years.

Without question, the development of muon-based accelerator facilities offers great scientific promise and remains a worthy—albeit challenging—goal to pursue.

## Acknowledgments

This work was supported by the Director, Office of Science, Office of High Energy Physics, of the U. S. Dept. of Energy, under Contract No. DE-AC02-05CH11231. I thank all my colleagues in MAP, MICE, and the IDS-NF for their dedication and skill in carrying out the work described here.